\documentclass[aps,prb,showpacs,nofootinbib,twocolumn,floatfix,superscriptaddress]{revtex4}
\pdfoutput=1
\usepackage{latexsym} \usepackage{amsmath} \usepackage{dcolumn}
\usepackage{bm} \usepackage{graphicx} \usepackage{epsfig}
\usepackage{float}
\usepackage{url}

\usepackage[]{graphicx}
\usepackage[]{wrapfig}
\usepackage[]{color}
\begin{document}

\title{
 {\textit {\textbf{GW}}} quasiparticle calculations with spin-orbit coupling for the light actinides
}
\author{Towfiq Ahmed}
\affiliation{
Theoretical Division, Los Alamos
National Laboratory, Los Alamos, New Mexico 87545}

\author{R. C. Albers}
\affiliation{
Theoretical Division, Los Alamos
National Laboratory, Los Alamos, New Mexico 87545}

\author{A. V. Balatsky}
\affiliation{
Theoretical Division, Los Alamos
National Laboratory, Los Alamos, New Mexico 87545}
\affiliation{Center for Integrated Nanotechnologies, Los Alamos
National Laboratory, Los Alamos, New Mexico 87545}
\affiliation{NORDITA, KTH Royal Institute of Technology and Stockholm University,
Roslagstullsbacken 23, 106 91 Stockholm
Sweden}

\author{C. Friedrich}
\affiliation{Peter Gr{\"u}nberg Institut and
Institute for
Advanced Simulation, Forschungszentrum J{\"u}lich and JARA, D-52425 J{\"u}lich, Germany}

\author{Jian-Xin Zhu}
\email[To whom correspondence should be addressed. \\ Electronic address: ]{jxzhu@lanl.gov}
\homepage{http://theory.lanl.gov}
\affiliation{
Theoretical Division, Los Alamos
National Laboratory, Los Alamos, New Mexico 87545}
\affiliation{Center for Integrated Nanotechnologies, Los Alamos
National Laboratory, Los Alamos, New Mexico 87545}


\date{\today}



\begin{abstract}
We report on the importance of $GW$ self-energy corrections for the
electronic structure of
light actinides in the weak-to-intermediate coupling regime. Our study is
based on calculations of the band structure and total density of states of Np, U, and Pu
using a one-shot $GW$ approximation 
that includes spin-orbit coupling within a full potential LAPW framework.
We also present RPA screened effective Coulomb interactions for the $f$-electron orbitals for different lattice constants, and show that there is an
increased contribution from electron-electron correlation 
in these systems for expanded lattices.
We find a significant amount of
electronic correlation in these highly localized electronic systems.
\end{abstract}


\maketitle
\section{Introduction} With significantly localized and partially filled $f$-electrons, the light actinide metals
have both strong electronic correlation effects and spin-orbit (SO) coupling present in their electronic structure.
Many theoretical tools have been developed in the recent years in order to address the strong correlation aspect, which is still considered to be one of the most challenging problems of
modern condensed matter physics.
For example, many-body treatments of the model Hamiltonian approach, such as Hubbard~\cite{hubbard} and periodic Anderson~\cite{anderson} models, have been
extensively used to study and
explain the electronic structures of the narrow band systems. Particularly for the $\delta$-phase of plutonium (Pu), the dynamical mean-field theory (DMFT)~\cite{Kotliar_Nat} provides a theoretical volume in good agreement with the
experimental measurement.~\cite{Cooper2000}
The $\delta$-phase of Pu in particular involves a crossover of itinerant-to-localized behavior in the light
actinide series, and has hence been much studied. DMFT has been quite successful in
predicting its several electronic features, including the $5f$
occupancy of its valence band.~\cite{jianxin_1,jianxin_2,jianxin_3}

Within the scope of first-principles theory, several developments are currently in progress.
The LDA+$U$ method was first proposed by Anisimov
{\it {et al.}}~\cite{Anisimov97} Although Hubbard $U$ in this method is determined
parametrically, such hybrid
methods have been used successfully in accurate description of
electronic structure and spectroscopies of many systems
such as the transition-metal oxides and high-$T_c$ cuprates.~\cite{towfiq_2012,Patrick2009,Cococcioni_2005,Anisimov93,Nakamura_2006}
Addressing strong correlation in a completely parametric
free manner is often desirable, but requires going beyond the local density approximation (LDA) of conventional
density functional theory (DFT). Constrained random phase approximation (cRPA) and constrained LDA (cLDA) are
two most popular methods in estimating Hubbard parameters, albeit with limited
success.~\cite{Arya_2004,Arya_2006}
Such combinations have been further extended by constructing a quasi-particle $GW$
self-energy~\cite{hedin69} from single or multiband Hubbard
model~\cite{tanmoy} and
have been successfully implemented for calculating the spectroscopy of many
correlated $d$- or $f$-electron systems.~\cite{towfiq_2011,susmita}
For the light actinides and Pu, parameter-free $GW$ calculations~\cite{chantis} in the absence of SO coupling have shown significant band renormalization effects.

The second essential ingredient for understanding the electronic structure of the actinide elements is their strong  SO coupling, which must be incorporated simultaneously with the many-body correlation effects.
Within the $GW$ approximation, a Dirac-relativistic approach has been implemented in a fully self-consistent manner in order to study Pu and Am metals.~\cite{dirac-GW}
In this paper, spin-orbit coupling was implemented within a scalar relativistically 
framework that uses an 
$LS$ basis instead of the fully relativistic $JJ$ basis. The $LS$ scheme is particularly convenient
for most condensed matter systems.~\cite{cowan}  
In addition, Hund's rules have a simpler realization in an $LS$ basis 
when compared to a $JJ$ basis,\cite{cowan} 
and it is much easier to treat magnetism when spin and orbital quantum numbers can be 
clearly identified.  

In this paper, we have calculated the LDA and $GW$ renormalized band structure of
U, Np, Pu, and an extended Pu system. With increasing lattice constants and partially filled $f$-orbitals, these $5f$-electron systems allow us to
understand the correlation physics
from itinerant to localized behavior for elemental materials, where the
SO coupling is comparable to the effective 5$f$ band widths.
Within the same $GW$ approximation, we have also evaluated
 the
average screened Coulomb interaction $W(\omega=0)$.
In the weakly interacting electronic systems, the $GW$
self-energy is well known to incorporate the dynamic correlation
for both short- and long-ranged Coulomb interactions.
By using a standard first-principles method, our calculations provide an important benchmark of the
significance of the
correlation strength in the light actinides that might further be refined
using $GW$ as a starting point while also including the effects of SO coupling.
\section{Formalism}
The relativistic extension of the quasiparticle
correction
is straight forward due to the single-particle nature of spin-orbit (SO) 
interaction term, which can be simply added to the single-particle
Hamiltonian.
Because of the coupling between the spin and orbital degrees of freedom,
the projected spin operator $\hat{S_z}$ is no longer a good quantum
number. Therefore, diagonalization of the Hamiltonian gives
Bloch states, which
can be written as a linear combination of both spin up ($\uparrow$) and
down ($\downarrow$) states,
\begin{equation}
\psi_{\bf{k}n}(\bf{r},s)=\psi_{\bf{k}n}^{\uparrow}(\bf{r})\chi^{\uparrow}(s)+
\psi_{\bf{k}n}^{\downarrow}(\bf{r})\chi^{\downarrow}(s)\;.
\end{equation}
Here, {\bf k} is the Block vector, {\bf s} is the spin degrees of freedom, 
 and {\bf n} is band index. 

Accordingly, the single-particle Green's function exhibits 
off-diagonal elements in spin space (up-down an down-up) 
enabling spin-flip processes. 
The relativistic generalization~\cite{general1,general2} of Hedin's $GW$ equations~\cite{hedin69,Hedin_89}
thus begins
with a spin-dependent formulation of the Green's function,
\begin{equation}
G_{\alpha\beta}(\bf{r},\bf{r'};\omega)=\sum_{\bf{k}n}\frac{\psi_{\bf{k}n}^{\alpha}(\bf{r})\psi_{\bf{k}n}^{\beta \ast}(\bf{r'})}{\omega - \epsilon_{\bf{k}n} + 
{\it{i}} \eta sgn(\epsilon_{\bf{k}n}-\epsilon_F)}\;,
\end{equation}
where $\alpha$ and $\beta$ represent spin up ($\uparrow$) or down ($\downarrow$) states, $\eta$ is a positive infinitesimal, $\epsilon_{\bf{k}n}$ is the
eigenvalue for diagonalized single-particle Hamiltonian and $\epsilon_F$ is
the Fermi energy. Within the random phase approximation (RPA), the polarization
function can be obtained as
\begin{equation}
P(\bf{r},\bf{r'};\omega)=\frac{\it{-i}}{2\pi} \sum_{\alpha\beta} \int G_{\alpha\beta}(\bf{r},\bf{r'};\omega+\omega')G_{\beta\alpha}(\bf{r'},\bf{r};\omega')d\omega'.
\end{equation}
The screened Coulomb interaction $W=V+VPW$ therefore indirectly depends on
the SO coupling through the spin-dependent Green's function. Here $V$ is
the bare unscreened Coulomb interaction before the RPA screening, and $W$ is 
RPA screened Coulomb interaction, $W(\bf{r},\bf{r'},\omega) = \epsilon_{RPA}^{-1}(\bf{r}, \bf{r'},\omega) V(\bf{r},\bf{r'})$.
Finally the spin-dependent matrix elements of the $GW$ self-energy can
be constructed following Hedin's prescription as
\begin{equation}
\Sigma_{\alpha\beta}(\bf{r},\bf{r'};\omega)=\frac{\it{i}}{2\pi} \int G_{\alpha\beta}(\bf{r},\bf{r'};\omega+\omega')W(\bf{r},\bf{r'};\omega')e^{\it{i} \eta \omega'}d\omega'. 
\end{equation}
A one-shot $GW$ approach is equivalent to the leading order
perturbation, and the quasi-particle eigenfunctions are generally
approximated as
the Bloch functions $\psi_{\bf{k}n}^{\uparrow / \downarrow}$.
The quasi-particle correction for the eigenvalues is then
\begin{equation}
E_{\bf{k}n}=\epsilon_{\bf{k}n}+\sum_{\alpha\beta} \langle \psi_{\bf{k}n}^{\alpha}|\Sigma_{\alpha\beta}(E_{\bf{k}n})-v^{xc} \delta_{\alpha\beta} |\psi_{\bf{k}n}^{\beta}\rangle\;.
\end{equation}
All of the above equations are a spin-dependent generalization for
the $GW$ approximation. 


To evaluate the effect of lattice spacing and localization of the $f$-orbitals on 
the correlation strength, we have calculated $V_{eff}$ and $U_{eff}$ as shown in Fig.~4, 
where $V_{eff}$ and $U_{eff}$ are the average onsite bare and 
RPA-screened Coulomb interactions of the localized $f$ electrons. These quantities 
are calculated using the Anisimov prescription~\cite{Anisimov97} by averaging over
the orbitals ($m_1,m_2,m_3,$ and $m_4$) of the angular-momentum-projected bare and 
screened Coulomb interactions, which are correspondingly defined as 
$\langle l_1,m_1;l_2,m_2|V|l_3,m_3;l_4,m_4\rangle$ and   
$\langle l_1,m_1;l_2,m_2|W|l_3,m_3;l_4,m_4 \rangle$. 
To compare with the experimental photoemission spectrum, the spectral function 
is calculated from the convolution between the density of states (DOS) 
multiplied by the fermi function 
and a Lorentzian function. 
The purpose of this convolution is to account for the experimental broadening.~\cite{shorikov} 
It is  written as 
\begin{equation}
A(\omega)=\int_{-\infty}^{\infty} \rho(\omega') \frac{\sigma}{(2\pi(\omega-\omega'))^2+(\sigma/2)^2} \frac{1}{e^{\omega'/kT}+1} d\omega',  
\end{equation} 
where $\rho(\omega)$ is the total DOS, $k$ is the Boltzman constant, 
$T$ is temperature, at which the experiment was performed, 
and $\sigma$ is the experimental broadening. 

\section{Computational Methodology} For most of our calculations, 
we have used the one-shot relativistic extension~\cite{spex_so-gw} of the
$GW$ self-energy correction implemented in the code SPEX.~\cite{spex_main} 
In this approach, when the SO coupling is present, 
the single-particle Green's function is represented 
in terms of the spin-dependent Bloch band states. 
The polarization function is then evaluated within the RPA approximation, 
which determines the screened Coulomb interaction and single-particle self energy.  
The latter gives the quasiparticle correction. 
We refer to 
Ref.~\onlinecite{spex_main} for more technical details.

In practice, computational methods are developed by
expressing the Bloch states in a suitable basis such as plane wave, linear muffin-tin orbital or augmented plane wave.
On the Kohn-Sham level, the SO coupling term is incorporated in a second-variational step,~\cite{freeman}
where the single-particle Hamiltonian including SO coupling is diagonalized in the basis of single-particle states
that are eigenfunctions of the Hamiltonian without the SO term. 
Therefore, the SO coupling effect on the
quasiparticle correction is naturally included through the Kohn-Sham single-particle states,
without the need for an {\it {a posteriori}} treatment, see Ref.~\onlinecite{aguilera} for a
detailed discussion.

In our calculations we used
a full potential LAPW basis with a SO interaction in the DFT code
FLEUR~\cite{fleur} along with a quasi-particle correction in the
one-shot $GW$ code, SPEX.~\cite{spex_main}
\begin{figure}
 \includegraphics[scale=0.34,angle=0]{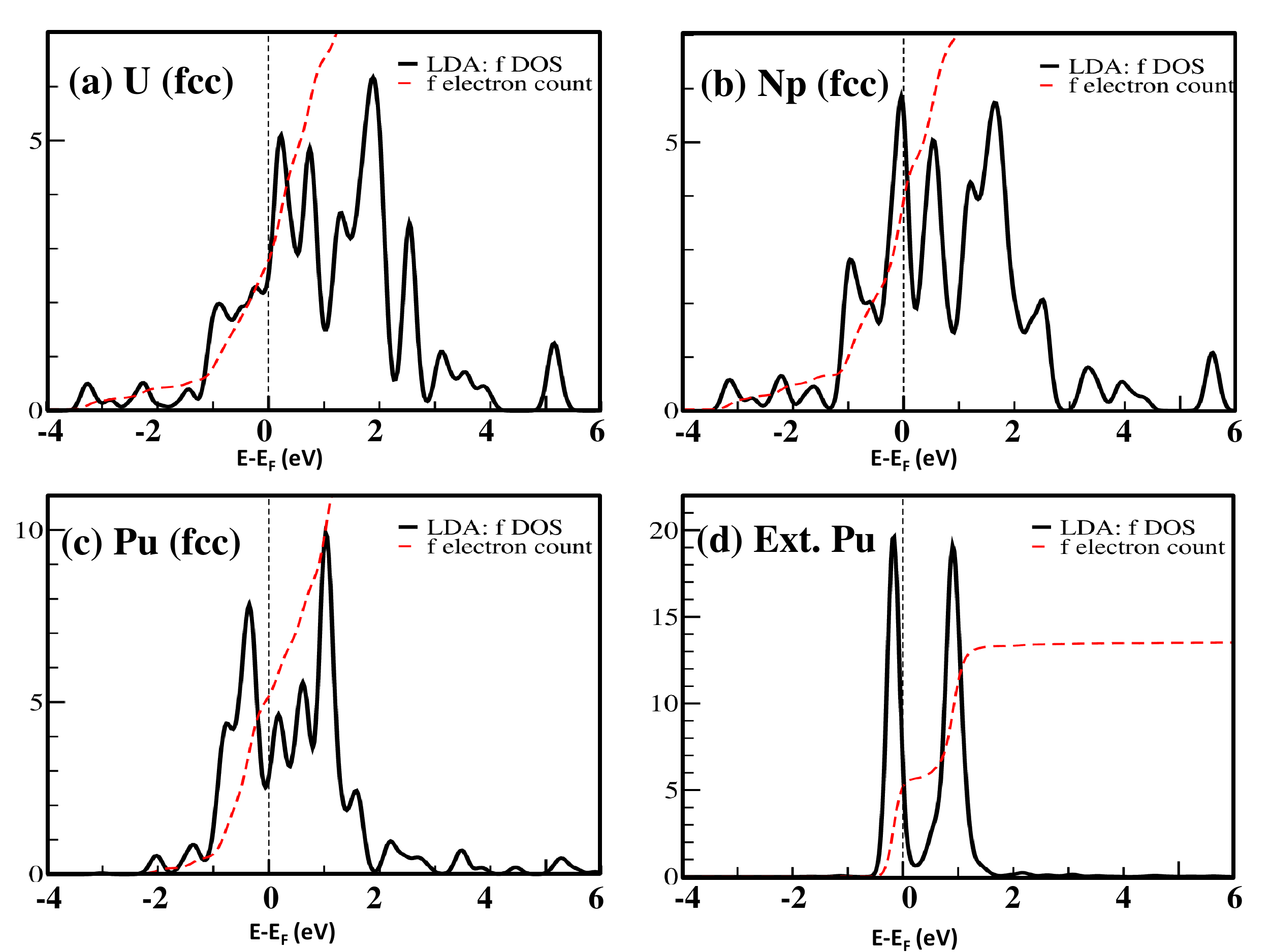}
  \caption
   {(Color online) LDA calculated angular momentum projected $f$-DOS for fcc U, Np, $\delta$-Pu and
extended Pu. The vertical dashed line is at the Fermi energy, E$_F$ = 0 eV.
The dashed curve (red) is the cumulative valence $f$-electron number (integrated $f$-DOS).
Spin-orbit splitting is most noticeable for the heavier (more localized) actinides.
}\label{fig1}
\end{figure}
Unlike other $GW$ relativistic calculations,~\cite{dirac-GW} instead
of an imaginary frequency approach that is analytically continued to
real frequencies, our method employs a contour-integration technique in which the self-energy
is calculated directly on a real-frequency mesh.
This avoids all the well known sensitivities of the results 
on the analytic-continuation algorithms,~\cite{gunnarsson2010}
and hence reduces this uncertainty.
Our calculations are based on the one-shot $GW$ that involves a single iteration of the $GW$ equations, where the input quantities are the LDA eigenvalues and eigenfunctions.  
Our self-consistent LDA calculations were converged using 300 k-points in the irreducible 
Brillouin zone (IBZ), and the one-shot GW calculations with 64 k-points in the IBZ.
For metals, a one-shot GW approach has been shown to be in better agreement with experiment than fully self-consistent $GW$
calculations,~\cite{aryasetiawan_98,aulbur2000} 
which miss important vertex corrections that are believed to be essential 
for predicting the correct plasmon energies and for canceling other deficiencies
generated in the self-consistent cycle;
these defects seem to be absent in the one-shot approach.
Hence the standard one-shot calculation may provide a better benchmark than full self-consistency.

\section{Lattice structure and SO coupling}
Our investigation of electronic correlation from an intermediate to a strong-coupling
regime includes three light-actinide elements U, Np, and Pu.
This enables us to study variations 
in the correlation strength simultaneously with a changing SO coupling.
Within the phases of the light actinides, a high-temperature fcc $\delta$-Pu 
phase is by far the most interesting.
With a unit cell volume of 168 $a_0^3$ (with $a_0$ being the Bohr radius) 
and a lattice constant of $a_{Pu}=4.64 \AA$,
$\delta$-Pu is known to involve both itinerant and localized electrons, 
i.e., has a dual-nature $f$-orbital character.

In order to see clear trends in the physical and electronic properties, 
we have chosen to do all of our calculations in the same fcc crystal structure as $\delta$ Pu.
Other crystal structures would change relative near-neighbor distances and hence would
modify the various hybridizations present in the calculations in a way
that would modify correlation strength (for an example of how significant these changes
would be, see Ref.~\onlinecite{jianxin_2}).
This problem presents a quandary in that U and Np have no fcc phases.
To resolve this issue, we have chosen to do calculations for an fcc crystal structure for U and Np
at a volume per atom that is equivalent to the most reasonable 
high-temperature cubic phase in these elements,
which turns out to be bcc in both cases.  
More precisely, we refer to the $\gamma$-U and $\gamma$-Np bcc phases
with respective unit cell volumes~\cite{albers_volume} of 138.89 $a_0^3$ and
129.9 $a_0^3$.
We refer to our fcc calculations as $\gamma$-U (fcc) and $\gamma$-Np (fcc)
to indicate how we have chosen the lattice constant for these calculations, which have
the modified lattice constants, $a_{U}=4.35 \AA$ and
$a_{Np}=4.25 \AA$ that are both smaller than $a_{Pu}$ 
and hence are anticipated to show more itinerant
electronic behavior.
Finally, we have also considered an fcc-Pu system with an enlarged lattice
constant of $a=6.64 \AA$.
Such a fictitious system allows us to understand the
strong electronic correlation in an extremely localized limit, 
where the overlap and hybridization between the neighboring $5f$ orbitals is minimal.

Fig.~1 shows the LDA calculated angular-momentum-projected $f$-DOS 
for all of these cases.
For the enlarged Pu lattice constant calculations, 
as shown in Fig. 1(d), the dominant peak split near Fermi energy is mainly due to SO coupling 
and is approximately in the atomic limit.
In our LDA calculations for U and Np, 
hybridization, crystal-field, and SO splitting were about the same order 
of magnitude (Fig.~\ref{fig1}(a) and (b)).
With the slightly larger lattice constant in
$\delta$-Pu, the larger SO splitting separates the DOS into two peaks,
which are mainly $j=5/2$ for the lower peak and 7/2 for the upper peak, as
shown in Fig.~\ref{fig1}(c), whereas the two j-components are more mixed for U and Np.
Besides SO and crystal fields, which are well captured in DFT, one
must also consider the missing long- and short-ranged dynamical correlations,  which
are the main focus of the remaining sections.
For reference, some lattice parameters and calculated SO splitting from the $f$-DOS are listed in Table~\ref{table}.
\begin{figure}
 \includegraphics[scale=0.34,angle=0]{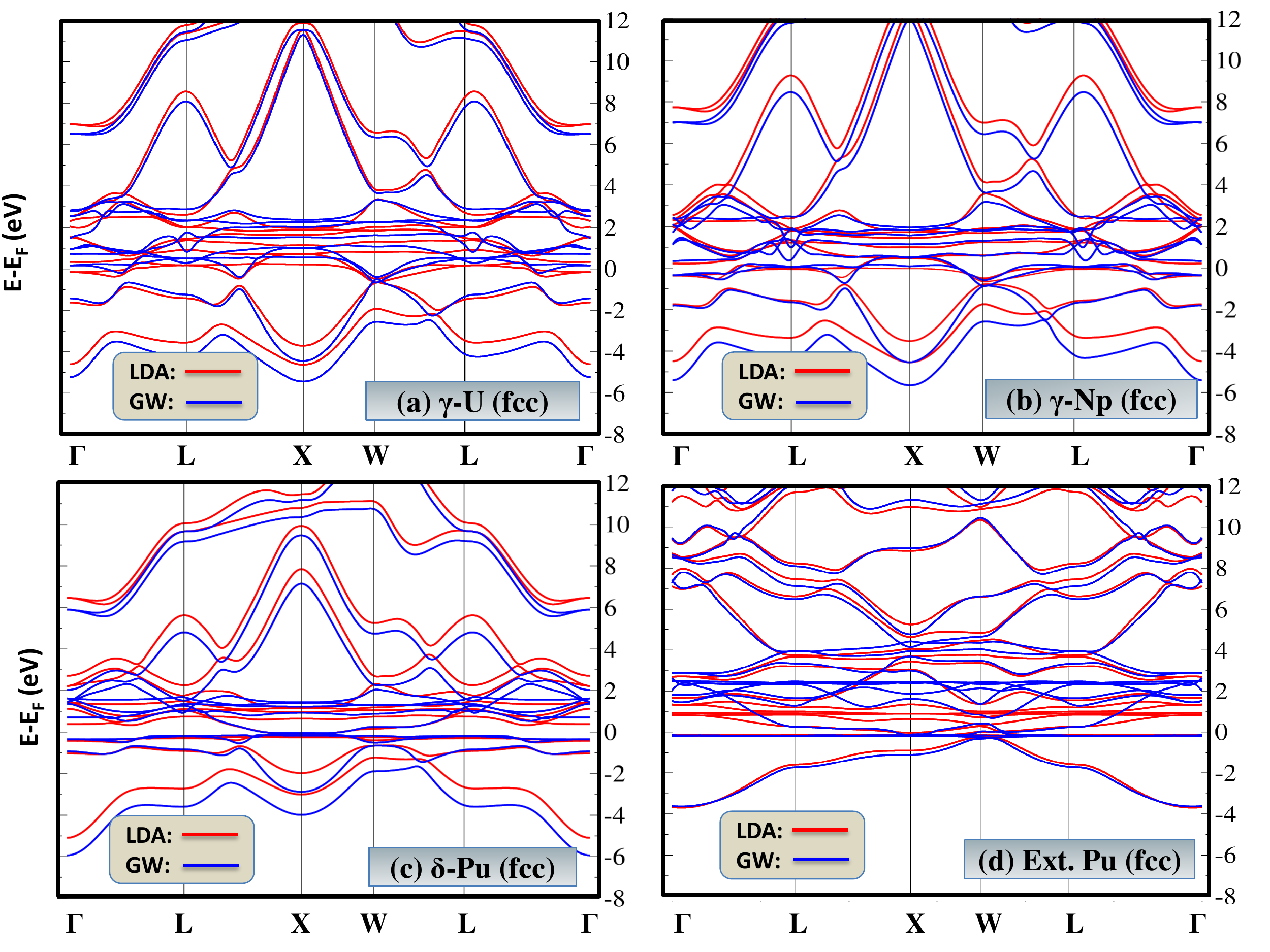}
  \caption
   {(Color online)
LDA (red) and $GW$ renormalized (blue) band structure of fcc U, Np, Pu and
extended Pu in (a),(b),(c), and (d) accordingly. Bands are calculated along
$\Gamma-L-X-W-L-\Gamma$ symmetry line and the Fermi energy is set at 0 eV.  An fcc crystal
structure is used for all calculations.
}\label{fig2}
\end{figure}
\begin{figure}
 \includegraphics[scale=0.34,angle=0]{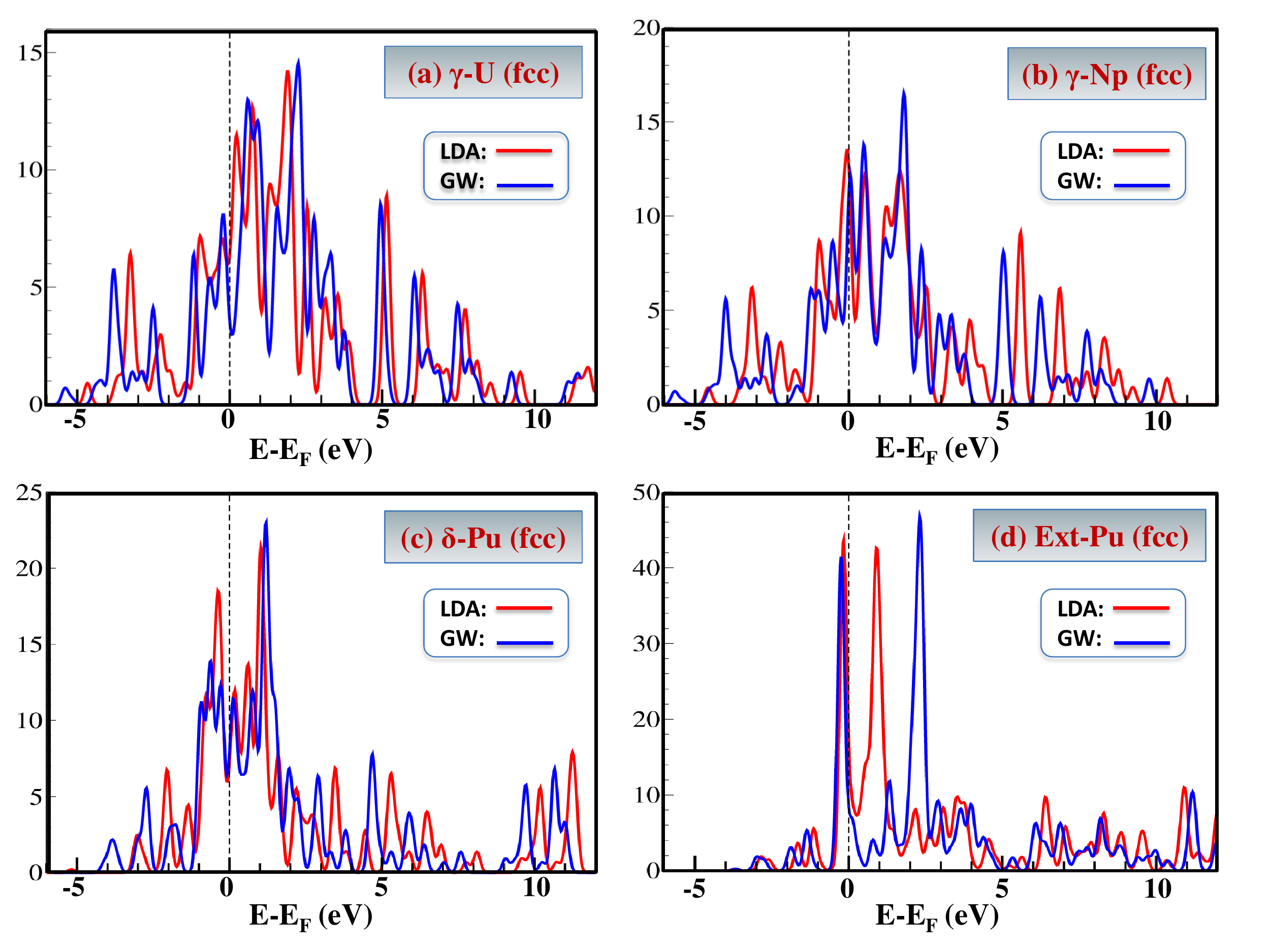}
  \caption
   {(Color online)
Total density of states corresponding to LDA (red) and $GW$ (blue) band-structure
calculations presented in Fig.2 for (a) U, (b) Np, (c) $\delta$-Pu, and (d) extended Pu.
Vertical dashed lines are at the Fermi energy at 0 eV.
}\label{fig3}
\end{figure}

\section{QP DOS and band renormalization}
In previous studies using a relativistic self-consistent $GW$ scheme 
for the actinide elements,~\cite{dirac-GW} the
importance of a quasi-particle treatment along with SO coupling was shown to be particularly
important for $\delta$-Pu.
In another $GW$ calculation without
SO coupling by Chantis {\it{et al.}},~\cite{chantis} the band structure of $\delta$-Pu was found
to have significant renormalization effect in the presence of crystal-fields, indicating
that electronic correlation of $\delta$-Pu lies in between intermediate and strong coupling limit.
Here we systematically investigate correlation in the electronic
structure for the four different systems including $\delta$-Pu
with different inter-atomic distances.
The real part of the self-energy $GW$ corrections causes a change in
the position of the Fermi energy due to the different size of the energy
shifts for the more localized $f$-states relative to the other itinerant
$s$-$p$-$d$ states.
To handle this,
in our metallic calculations we determine the Fermi energy according to its definition by
requiring that the integrated total DOS below the Fermi energy
have the correct number of electrons.
We then measure all energies with respect to the Fermi energy (E$_F=0$).

The bands near the Fermi energy are
plotted along $\Gamma$-L-X-W-L-$\Gamma$ high-symmetry line.
The $GW$ renormalized bands are plotted in blue lines while the LDA band
structures are shown in red lines in Fig.~\ref{fig2};
the same color scheme is followed for the total DOS shown in Fig.~\ref{fig3}.
As expected, fcc Np, which has the smallest lattice spacing $a_{Np}=4.25 \AA$, shows the largest dispersion of bands (Fig.~\ref{fig2}(b)).
On the other hand, the $5f$ electron bands in the extended $\delta$-Pu are the least dispersive.
The very localized extended Pu system has very flat bands (Fig.~\ref{fig2}(d)).
In this system, the $GW$ corrections cause a large shift in the unoccupied $j=7/2$ component of the $5f$ orbitals to higher energy (cf. Fig.~\ref{fig3}(d)) due to the non-local self-energy operator that acts like an
effective Hubbard $U$ in the calculations.
In general, the occupied valence electrons can be divided into itinerant states,
predominantly 7$s^2$ and 6$d^1$
(because of their high principal quantum numbers,
these have many nodes in the core region for orthogonalization
to lower principal quantum number atomic states,
which cause a large curvature or high kinetic energy), and localized states, the 5$f$ electrons.
Similar to $GW$ calculations on Uranium,~\cite{chantis2008}
the $f$ states are shifted up relative to the itinerant states
(bottom of the valence band).

\begin{table}[ht]
\caption{Experimental lattice parameters and calculated peak split ($\Delta$) and Coulomb interaction
for U, Np, Pu and extended Pu. }
\centering
\begin{tabular*}{0.50\textwidth}{@{\extracolsep{\fill}} l  c  c  c c }
\hline
Element & U & Np & Pu & ext-Pu\\[0.5ex]
\hline\hline \\[-0.3ex]
$Z$ & 92 & 93 & 94 & 94 \\
$5f$ valence occ. & 3 & 4 & 5 & 5 \\
Original crystal symm. & bcc ($\gamma$)& bcc ($\gamma$) & fcc ($\delta$)&fcc \\
Considered symm. & fcc & fcc & fcc & fcc \\
Unit-cell volume (a.u.) & 139.9 &129.9 & 168.0 & 560.0 \\
FCC lattice constants ($\AA$) & 4.35 & 4.25 & 4.64 & 6.64 \\
$\Delta_{SO}$ (eV) & 0.87 & 0.95 & 1.17 & 1.17 \\
$\Delta_{LDA}$ (eV) & 1.50& 1.69 & 1.35 & 1.17 \\
$\Delta_{GW}$ (eV) & 1.61 & 1.71 & 1.91 & 2.57 \\
$\Delta_{corr}=\Delta_{GW}-\Delta_{LDA}$ (eV) & 0.11 & 0.02 & 0.56 & 1.4 \\
$\Delta_{Xtal}=\Delta_{LDA}-\Delta_{SO}$ (eV) & 0.63 & 0.74 & 0.18 & 0.00 \\
$V_{init}$ (eV) & 8.23 & 4.47 & 10.21 & 10.36 \\
$W_{screened}$ (eV) & 5.68  & 2.36 & 7.74 & 8.69 \\
\hline
\end{tabular*}
\label{table}
\end{table}

The quasi-particle total DOS for all four systems are plotted (blue curve)
against LDA total DOS (red curve) in Fig.~\ref{fig3}.
By QP-DOS we mean that we take into account
only the shift in the quasiparticle energy due to the real-part of the $GW$
self-energy and ignore any life-time broadening from the imaginary part.
Although the QP-DOS is different from the true DOS, it is helpful for
identifying the peak locations and the effects of self-energy shifts
on them, which would otherwise be smeared out by the lifetime broadening.
\begin{figure}
 \includegraphics[scale=0.34,angle=0]{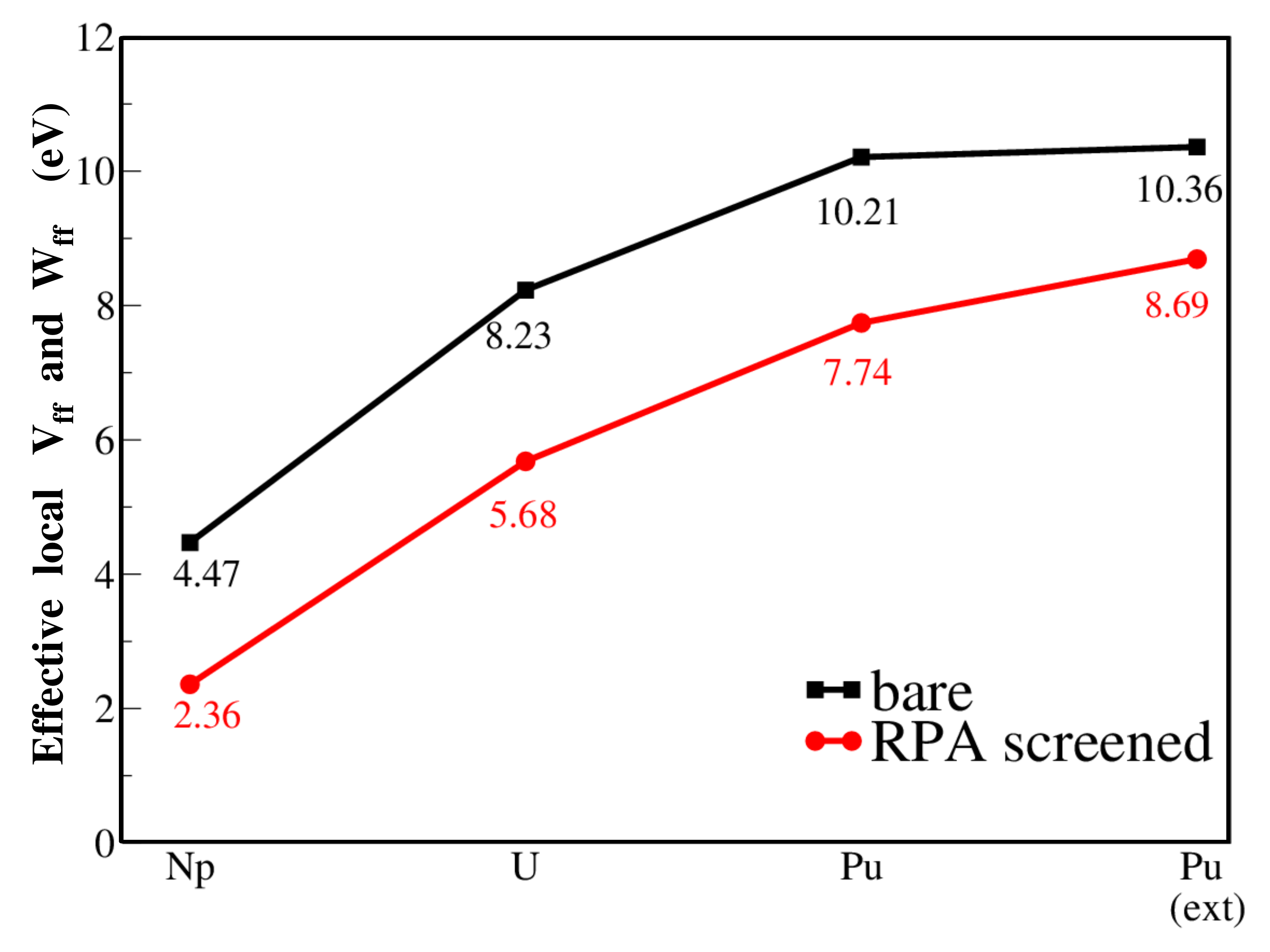}
  \caption
   {(Color online)
Comparison of experimental photoemission spectrum (dashed black line) of $\delta$-Pu with our calculations using $GW$+SO (solid blue), LDA+SO (solid red), 
$GW$ without SO (sashed blue) and LDA without SO (dashed red); Vertical dashed
lines show three peaks near E$_F$ on Photoemission spectra. Here T = 80 K and $\sigma$=0.2.   
}\label{fig4}
\end{figure}
In Table~\ref{table} we have calculated $\Delta_{SO}$ by using a very large
lattice spacing for U, Np and Pu similar to what is shown in Fig.~\ref{fig1}(d).
In this case hybridization is negligible and we are essentially in the atomic limit.
For $\Delta_{LDA}$ and
$\Delta_{GW}$ we calculated the distance between the dominant $5f$ peaks
near Fermi energy. Although the SO splitting is not as distinct due
to the crystal field effects, comparison between these peaks with and without
self-energy correction helps us reveal the contribution from dynamic correlations.
Qualitatively we can attribute $\Delta_{Xtal}$ (see Table~\ref{table}) as a measure of crystal field
effect relative to SO coupling, and $\Delta_{corr}$
as a measure of correlation correction.  From Table~\ref{table}, one finds
that the material with the smallest lattice constant, Np, has the most
itinerant behavior for the valence electrons with prominent
crystal field splitting and is least affected by the quasi-particle
correction. On the other hand, extended $\delta$-Pu shows the opposite trend.
The itinerant behavior is also evident from the presence of
several crystal-field splitting in the total DOS for both Np and U (Fig.~\ref{fig3}(a-b)).
\begin{figure}
 \includegraphics[scale=0.34,angle=0]{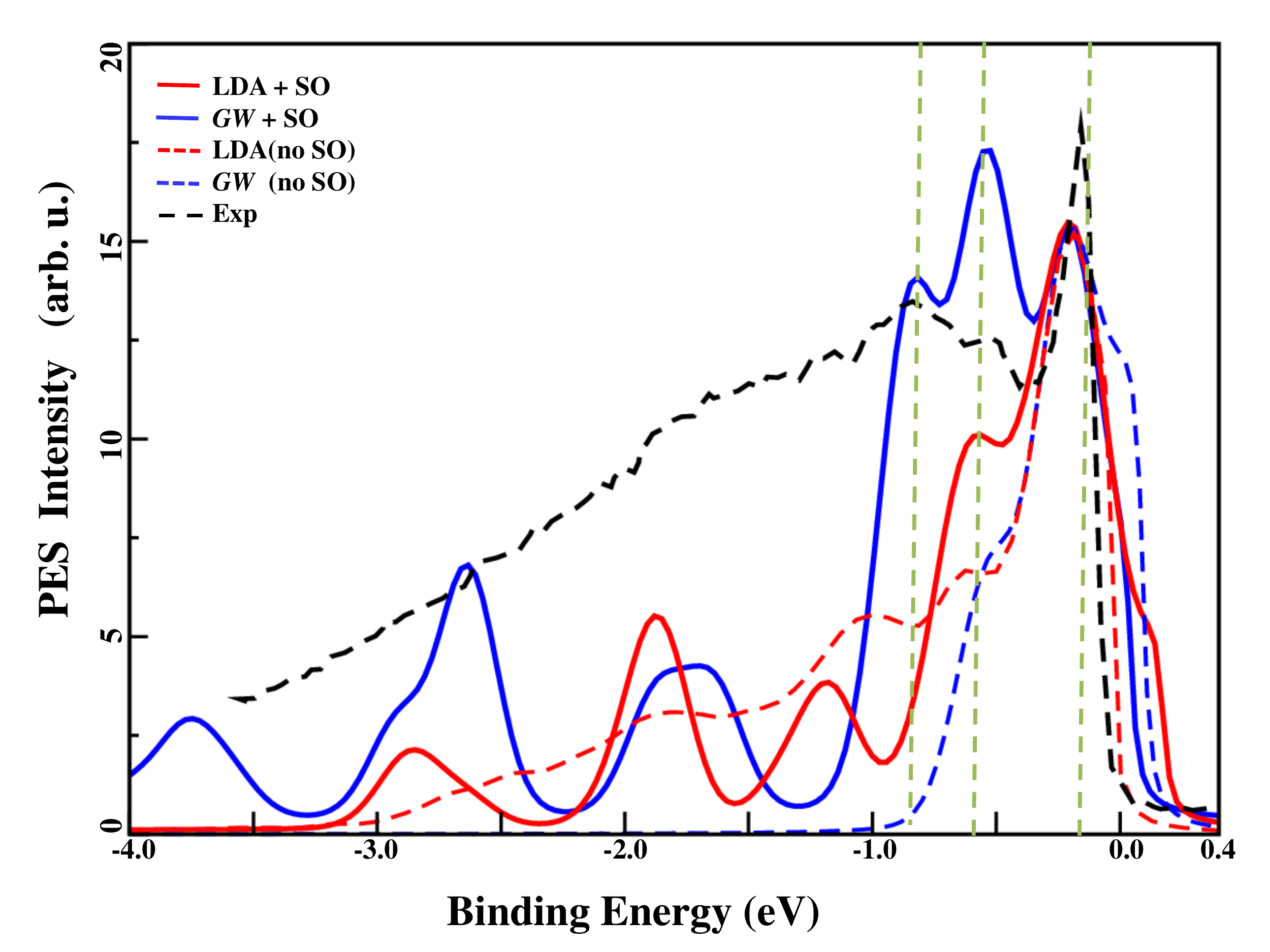}
  \caption
   {(Color online)
Earlier theoretical calculations of total DOS for $\delta$-Pu using 
fully relativistic GW (green) and scalar relativistic GW (magenta) method. 
Our 
calculations with LDA (solid red) and with one shot GW + SO approach (solid 
blue) are also shown. 
}\label{fig5}
\end{figure}
Spin-orbit splitting is more distinct for $\delta$-Pu in Fig.~\ref{fig3}(c), and the unoccupied $j$=7/2 peak at 1.2 eV shifts slightly to higher energy due to the $GW$ correction.
Other states that are between 2 and 12 eV shift significantly downward, and thus band narrowing is 
not only due to the $f$ like orbitals, but also involves other (e.g., 6d) electrons.
Similar findings have also been reported by other authors.~\cite{dirac-GW}
For extended Pu, where the inter-atomic distance is too large, the self-energy corrections
of the highly localized $5f$ electrons cause
the SO coupled peak at ~1.2 eV to
shift 1.5 eV further to the right (Fig.~\ref{fig3}(d)).

\begin{figure}
 \includegraphics[scale=0.34,angle=0]{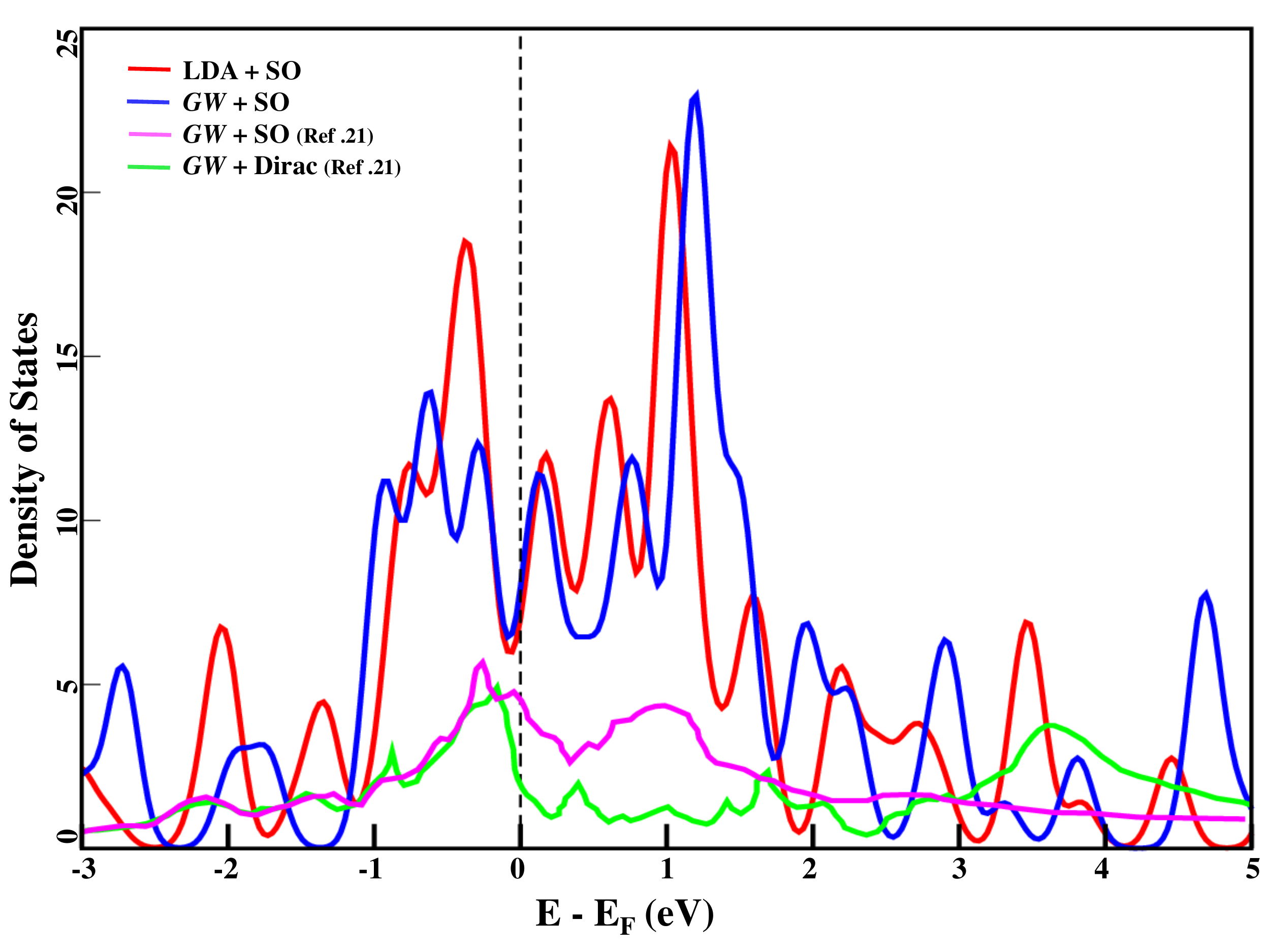}
  \caption
   {(Color online)
Average (effective) bare (black) and RPA screened (red) Coulomb interaction within occupied
$5f$ electrons for Np, U, Pu and extended Pu. The elements in {\it {x}}-axis are ordered
according to their increasing lattice spacing. The Coulomb interaction is
screened
using RPA response function. For more details see the text.
}\label{fig6}
\end{figure}

\section{Photoemission Spectra and DOS}
In Fig.~\ref{fig4} we have compared the available experimental data for $\delta$-Pu
with our calculated LDA and one-shot GW-based photoemission 
spectroscopy (PES) with and without the SO coupling. The band narrowing 
effect for the one-shot GW is evident for calculations without the SO coupling (see dashed blue 
line against dashed red line). The same effect was also obtained for elemental uranium solid within
the quasiparticle-GW without the SO coupling.~\cite{chantis2008} 
Inclusion of SO coupling does not show similar trend for GW calculations (solid blue) 
comparing to LDA (solid red). In the vicinity of Fermi energy, the band renormalization 
with one-shot GW+SO broadens the peak. Such results are also consistent with other theoretical 
calculations.\cite{dirac-GW}
Also, comparing with the experimental PES,~\cite{photo_exp} the GW+SO calculations (solid blue) are in 
better agreement with the three 
peak locations closest to the edge or Fermi energy, which are indicated by vertical 
dashed lines in Fig.~\ref{fig4}. 
Our real-frequency method for the self-energy is free from the 
sensitivities\cite{gunnarsson2010} caused by an analytical continuation approach. The conservation 
of occupied valence electrons and 
spectral weights are also consistent in our method, which can be noticed in Fig.~\ref{fig5} by 
comparing LDA (solid red) and GW (solid blue) calculations. On a similar scale, the GW+SO and GW+Dirac 
calculations as presented in Ref.~\onlinecite{dirac-GW} that use an analytical continuation method does not preserve the 
electron counts when compared 
to the LDA calculations. Although fully self-consistent and fully relativistic, such a method is not
completely free from these types of uncertainties.

\section{Effective localized Coulomb interaction}
With increasing lattice spacing, the partially filled $f$-orbitals become more localized and
onsite Coulomb interaction
becomes stronger. Such situations are most commonly realized as Hubbard like systems in model Hamiltonian approach,
where $U \gg t$ with $U$ being the Hubbard parameter (e.g. onsite Coulomb interaction),
and $t$ the hopping parameter. Within first-principles approach, there have been numerous
attempts to determine $U$ from
the electronic structure calculation such as the
constrained LDA (cLDA) and RPA (cRPA).~\cite{Arya_2004,Arya_2006}
The magnitude of $U$ often provides a good measure for static electronic correlation and used as an input
parameter for LDA+$U$ or LDA+DMFT calculations. 
Because our $GW$ calculations automatically include a screened Coulomb interaction 
$W(\omega)$ evaluated within an RPA response function,
it is useful to provide these results as another way to show a predicted correlation strength
for the different actinides.
Projecting $W$ on $f$-orbitals at $\omega=0$, in Fig.~\ref{fig6} we 
present the calculated local screened interaction $W_{ff}(\omega=0)$.
In addition, we also show the projected bare Coulomb interaction $V_{ff}$ for comparison. 
The elements are ordered according to their lattice constant along $x$-axis.
We observe that the screened Coulomb interaction scales with the bare one, both increasing with 
enlarged lattice spacing.

\section{Conclusion}
We have reported our findings on the electronic correlations in light actinide systems using
the one-shot $GW$ approximation with spin-orbit coupling included.
By systematically tuning from itinerant to localized regime in a set of $5f$ systems,
our calculations have shown the effectiveness of relativistic $GW$ correction
in describing the correct behavior of electronic structure in the intermediate coupling regime.
Thus, our calculations provide an important benchmark on the way to a complete
description of the electronic structure of light actinides that might be
further refined within the $GW$+DMFT~\cite{Biermann2003} like methods using $GW$ as a starting
point.

\section{Acknowledgement}
We thank A. Svane, N. E. Christensen, M. van Schilfgaarde, and A. N. Chantis for useful discussions and collaboration on related work.  This work was supported by U.S. DOE  at
LANL  under Contract No. DE-AC52-06NA25396,  the LANL LDRD Program
(T.A., R.C.A. \& J.-X.Z.),  and the Europe VR Program (A.V.B.).


\end{document}